\documentstyle[12pt]{article}
\begin{document}

\begin{titlepage}

\begin{flushright}
IFT -- P.075/97 \\
\end{flushright}

\vspace{1cm}
\begin{center}
{\Large  \bf
Minimal Supersymmetric Standard Model \\[3mm]
within CompHEP software package}

\vspace{15mm}
{\bf A.S.Belyaev} $^{\mbox{a,b,}}$\footnote{e-mail:
belyaev@monet.npi.msu.su},
{\bf A.V.Gladyshev} $^{\mbox{c,}}$\footnote{e-mail:
gladysh@thsun1.jinr.ru},
{\bf A.V.Semenov} $^{\mbox{d,}}$\footnote{e-mail:
semenov@theory.npi.msu.su}

\vspace{10mm}

$^{\mbox{a}}$ {\it Instituto de F\'\i sica Te\' orica, Universidade 
Estadual Paulista,\\
Rua Pamplona 145, 01405-900 - S\~ao Paolo, S.P., Brasil}

\vspace{5mm}

$^{\mbox{b}}$ {\it Skobeltsin Institute for Nuclear Physics,
Moscow State University, \\ 119 899, Moscow, Russian Federation}

\vspace{5mm}

$^{\mbox{c}}$ {\it Bogoliubov Laboratory of Theoretical Physics,
     Joint Institute for Nuclear Research, \\
	141 980 Dubna, Moscow Region, Russian Federation}

\vspace{5mm}

$^{\mbox{d}}$ {\it Laboratory of Particle Physics,
     Joint Institute for Nuclear Research, \\
	141 980 Dubna, Moscow Region, Russian Federation}
\end{center}

\vspace{10mm}

\begin{abstract}
The Minimal Supersymmetric Standard Model is presented as a model 
for the CompHEP software package as a set of files containing 
the complete Lagrangian of the MSSM, particle contents and parameters.
All resources of CompHEP with a user-friendly interface are now 
available for the phenomenological study of the MSSM. Various special 
features of the model are discussed.
\end{abstract}

 \vskip 0.5cm
 
 \centerline{\sc Submitted to Comp. Phys. Commun.}
 \vskip 0.4cm
\end{titlepage}

%
%
\section{Introduction}

Supersymmetry (SUSY) is one of the most promising theoretical ideas
pretending to solve some problems of the Standard Model (SM) and Grand
Unified Theories (GUT). The simplest supersymmetric extension of the SM
is the Minimal Supersymmetric Standard Model (MSSM). The
phenomenological study of the MSSM in relation to colliders of TeV
energies is an important task for understanding the SUSY discovery
potential of existing and forthcoming accelerators.

One of the important points for the phenomenological study of differnt 
models of particle physics is the possibility for automatic calculations. 
There are several software packages (GRACE~\cite{grace}, DILL~\cite{dill}, 
XLOOPS~\cite{xloops}, PHYSICA~\cite{physica}, SHELL2~\cite{shell2}, 
TLAMM~\cite{tlamm}, HECAS~\cite{hecas}, etc.) which allow users to perform 
analytic and numerical calculations of high energy physics processes. 
Among them is the CompHEP software package~\cite{comphep} which was 
developed by High Energy Physics Group of Skobeltsin Institute for Nuclear 
Physics of Moscow State University. It allows to pass automatically from 
the Lagarngian of a model to event distribution and performs complete 
tree level calculations in the framework of any fed model. One of the main 
advantages of CompHEP is that it allows to perform the calculations within 
any user defined model. In comparison with the previous packages we present 
the {\it complete} version of the MSSM as a CompHEP model. Some technical 
details of implementing the MSSM into the package are discussed as well.

Files containing the MSSM Lagrangian, particles and parameters are available 
from the following WWW sites: 
\par
{\sl http://theory.npi.msu.su/\~{}comphep/comphep-susy.Z} or 
\par
{\sl http://thsun1.jinr.ru/\~{}comphep/comphep-susy.Z}.

%
%
\section{MSSM in the CompHEP package:
Lagrangian, particles and parameters}

The general structure of the CompHEP software package as well as its main 
features is described in Ref.~\cite{comphep}, and we will not repeat it here.
We just note that in order to specify a model for performing calculations
within CompHEP, one has to create the following set of files (this can be done 
aslo within the CompHEP package itself by choosing the menu item 
{\sf NEW MODEL}):

\bigskip
\begin{tabular}{lp{110mm}}
\verb+ lgrng+$N$\verb+.mdl + &  the table contatning the Feynman rules of 
the model \\
\verb+ prtcls+$N$\verb+.mdl + & the table with the particles of the model \\
\verb+ vars+$N$\verb+.mdl + &   the table of the model parameters \\
\verb+ func+$N$\verb+.mdl + &   the table with parameter dependences, 
which alows to have only independent parameters for the final calculation \\
\end{tabular}

\bigskip
\noindent
Here $N$ is an integer number, the number of the model.

We skip here the description of constructing the supersymmetric extension 
of the SM, since a number of exellent reviews are available, see 
{\it e.g.}~\cite{nilles,kane}. The Yukawa intreractions are determined
by the superpotential which in the case of the MSSM reads
$$
W = \epsilon_{ij}\left( h_L^{IJ} L^j_I H_1^i E^c_J 
	              + h_D^{IJ} Q^j_I H_1^i D^c_J
	              + h_U^{IJ} Q^j_I H_2^i U^c_J  
                      + \mu H_1^i H_2^j \right).
$$
Here $h^{IJ}$ are the Yukawa coupling constants, $L$ and $Q$ are the $SU(2)$ 
doublet lepton and quark superfields, $E^c$, $U^c$ and $D^c$ are the $SU(2)$ 
singlet charge-conjugated superfields of leptons and up- and down-type quarks,
$H_{1,2}$ are the $SU(2)$ doublet Higgs superfields, $i,j$ are the $SU(2)$
indices and $I,J$ are generation indices.

We follow the Feynman rules of the MSSM according to
the Ref.~\cite{rosiek}. Besides we have generated the Feynman rules by
ourselves using the general form of the Lagrangian~\cite{rosiek} for double 
check. It has been done by means of the LanHEP program~\cite{andrey}. 

The MSSM particle spectrum is the following:
\begin{itemize}
\item Standard Model particles:
\begin{itemize}
\item photon $\gamma$, $W^\pm$ and $Z$ bosons and gluon $g$
\item quarks $u$, $d$, $s$, $c$, $b$, $t$
\item leptons $e$, $\nu_e$, $\mu$, $\nu_\mu$, $\tau$, $\nu_\tau$
\end{itemize}
\item Higgs bosons $H_1$ and $H_2$ (the physical states are two neutral
$CP$-even Higgses $h$ and $H$, neutral $CP$-odd Higgs $A$ and a pair of
charged Higgs bosons $H^\pm$)
\item Superpartners 
\begin{itemize}
\item superpartners of the gauge bosons (gaugino) and Higgs bosons
(higgsino) whose mass eigenstates are two charginos
$\widetilde\chi^\pm_1$, $\widetilde\chi^\pm_2$ and four neutralinos
$\widetilde\chi^0_1$, $\widetilde\chi^0_2$, $\widetilde\chi^0_3$,
$\widetilde\chi^0_4$
\item superpartners of the matter fields (squarks and sleptons); the
physical eigenstates are mixtures of the superpartners of the
left-handed and right-handed quarks and leptons.
\end{itemize}
\end{itemize}

The part of file \verb+prtcls+$N$\verb+.mdl+ is presented in Table~1 
in the CompHEP notations. It contains all the particles mentioned above. 
The rest part of \verb+prtcls+$N$\verb+.mdl+ contains auxiliary fields
and is discussed later.

{\small
\begin{table}[htb]
\vspace*{-20mm}
\begin{center}
\begin{tabular}{ l | l | l | l | l | l | l | l }
\verb+ Full  name  + & \verb+ P + & \verb+ aP+ & \verb+2*spin+ & \verb+
mass + & \verb+width + & \verb+color+ & \verb+aux+ \\
\verb+photon       + & \verb+A  + & \verb+A  + & \verb+2     + &
\verb+0     + & \verb+0     + & \verb+1    + & \verb+G+ \\
\verb+Z boson      + & \verb+Z  + & \verb+Z  + & \verb+2     + &
\verb+MZ    + & \verb+wZ    + & \verb+1    + & \verb+G+ \\
\verb+gluon        + & \verb+G  + & \verb+G  + & \verb+2     + &
\verb+0     + & \verb+0     + & \verb+8    + & \verb+G+ \\ 
\verb+W boson      + & \verb=W+ = & \verb+W- + & \verb+2     + &
\verb+MW    + & \verb+wW    + & \verb+1    + & \verb+G+ \\
\verb+neutrino     + & \verb+n1 + & \verb+N1 + & \verb+1     + &
\verb+0     + & \verb+0     + & \verb+1    + & \verb+L+ \\ 
\verb+electron     + & \verb+e1 + & \verb+E1 + & \verb+1     + &
\verb+0     + & \verb+0     + & \verb+1    + & \\
\verb+mu-neutrino  + & \verb+n2 + & \verb+N2 + & \verb+1     + &
\verb+0     + & \verb+0     + & \verb+1    + & \verb+L+ \\
\verb+muon         + & \verb+e2 + & \verb+E2 + & \verb+1     + &
\verb+Mm    + & \verb+0     + & \verb+1    + & \\
\verb+tau-neutrino + & \verb+n3 + & \verb+N3 + & \verb+1     + &
\verb+0     + & \verb+0     + & \verb+1    + & \verb+L+ \\
\verb+tau-lepton   + & \verb+e3 + & \verb+E3 + & \verb+1     + &
\verb+Mt    + & \verb+0     + & \verb+1    + & \\
\verb+u-quark      + & \verb+u  + & \verb+U  + & \verb+1     + &
\verb+0     + & \verb+0     + & \verb+3    + & \\
\verb+d-quark      + & \verb+d  + & \verb+D  + & \verb+1     + &
\verb+0     + & \verb+0     + & \verb+3    + & \\
\verb+c-quark      + & \verb+c  + & \verb+C  + & \verb+1     + &
\verb+Mc    + & \verb+0     + & \verb+3    + & \\
\verb+s-quark      + & \verb+s  + & \verb+S  + & \verb+1     + &
\verb+Ms    + & \verb+0     + & \verb+3    + & \\
\verb+t-quark      + & \verb+t  + & \verb+T  + & \verb+1     + &
\verb+Mtop  + & \verb+wtop  + & \verb+3    + & \\
\verb+b-quark      + & \verb+b  + & \verb+B  + & \verb+1     + &
\verb+Mb    + & \verb+0     + & \verb+3    + & \\
\verb+Light Higgs  + & \verb+h  + & \verb+h  + & \verb+0     + &
\verb+Mh    + & \verb+wh    + & \verb+1    + & \\
\verb+Heavy higgs  + & \verb+H  + & \verb+H  + & \verb+0     + &
\verb+MHH   + & \verb+wHh   + & \verb+1    + & \\
\verb+3rd Higgs    + & \verb+H3 + & \verb+H3 + & \verb+0     + &
\verb+MH3   + & \verb+wH3   + & \verb+1    + & \\
\verb+Charged Higgs+ & \verb=H+ = & \verb+H- + & \verb+0     + &
\verb+MHc   + & \verb+wHc   + & \verb+1    + & \\
\verb+chargino 1   + & \verb=~1+= & \verb+~1-+ & \verb+1     + &
\verb+MC1   + & \verb+wC1   + & \verb+1    + & \\
\verb+chargino 2   + & \verb=~2+= & \verb+~2-+ & \verb+1     + &
\verb+MC2   + & \verb+wC2   + & \verb+1    + & \\
\verb+neutralino 1 + & \verb+~o1+ & \verb+~o1+ & \verb+1     + &
\verb+MNE1  + & \verb+wNE1  + & \verb+1    + & \\
\verb+neutralino 2 + & \verb+~o2+ & \verb+~o2+ & \verb+1     + &
\verb+MNE2  + & \verb+wNE2  + & \verb+1    + & \\
\verb+neutralino 3 + & \verb+~o3+ & \verb+~o3+ & \verb+1     + &
\verb+MNE3  + & \verb+wNE3  + & \verb+1    + & \\
\verb+neutralino 4 + & \verb+~o4+ & \verb+~o4+ & \verb+1     + &
\verb+MNE4  + & \verb+wNE4  + & \verb+1    + & \\
\verb+gluino       + & \verb+~g + & \verb+~g + & \verb+1     + &
\verb+MSG   + & \verb+wSG   + & \verb+8    + & \\
\verb+1st selectron+ & \verb+~e1+ & \verb+~E1+ & \verb+0     + &
\verb+MSe1  + & \verb+wSe1  + & \verb+1    + & \\
\verb+2nd selectron+ & \verb+~e4+ & \verb+~E4+ & \verb+0     + &
\verb+MSe2  + & \verb+wSe2  + & \verb+1    + & \\
\verb+1st smuon    + & \verb+~e2+ & \verb+~E2+ & \verb+0     + &
\verb+MSmu1 + & \verb+wSmu1 + & \verb+1    + & \\
\verb+2nd smuon    + & \verb+~e5+ & \verb+~E5+ & \verb+0     + &
\verb+MSmu2 + & \verb+wSmu2 + & \verb+1    + & \\
\verb+1st stau     + & \verb+~e3+ & \verb+~E3+ & \verb+0     + &
\verb+MStau1+ & \verb+wStau1+ & \verb+1    + & \\
\verb+2nd stau     + & \verb+~e6+ & \verb+~E6+ & \verb+0     + &
\verb+MStau2+ & \verb+wStau2+ & \verb+1    + & \\
\verb+e-sneutrino  + & \verb+~n1+ & \verb+~N1+ & \verb+0     + &
\verb+MSne  + & \verb+wSne  + & \verb+1    + & \\
\verb+m-sneutrino  + & \verb+~n2+ & \verb+~N2+ & \verb+0     + &
\verb+MSnmu + & \verb+wSnmu + & \verb+1    + & \\
\verb+t-sneutrino  + & \verb+~n3+ & \verb+~N3+ & \verb+0     + &
\verb+MSntau+ & \verb+wSntau+ & \verb+1    + & \\
\verb+u-squark 1   + & \verb+~u1+ & \verb+~U1+ & \verb+0     + &
\verb+MSu1  + & \verb+wSu1  + & \verb+3    + & \\
\verb+u-squark 2   + & \verb+~u2+ & \verb+~U2+ & \verb+0     + &
\verb+MSu2  + & \verb+wSu2  + & \verb+3    + & \\
\verb+d-squark 1   + & \verb+~d1+ & \verb+~D1+ & \verb+0     + &
\verb+MSd1  + & \verb+wSd1  + & \verb+3    + & \\
\verb+d-squark 2   + & \verb+~d2+ & \verb+~D2+ & \verb+0     + &
\verb+MSd2  + & \verb+wSd2  + & \verb+3    + & \\
\verb+c-squark 1   + & \verb+~c1+ & \verb+~C1+ & \verb+0     + &
\verb+MSc1  + & \verb+wSc1  + & \verb+3    + & \\
\verb+c-squark 2   + & \verb+~c2+ & \verb+~C2+ & \verb+0     + &
\verb+MSc2  + & \verb+wSc2  + & \verb+3    + & \\
\verb+s-squark 1   + & \verb+~s1+ & \verb+~S1+ & \verb+0     + &
\verb+MSs1  + & \verb+wSs1  + & \verb+3    + & \\
\verb+s-squark 2   + & \verb+~s2+ & \verb+~S2+ & \verb+0     + &
\verb+MSs2  + & \verb+wSs2  + & \verb+3    + & \\
\verb+t-squark 1   + & \verb+~t1+ & \verb+~T1+ & \verb+0     + &
\verb+MStop1+ & \verb+wStop1+ & \verb+3    + & \\
\verb+t-squark 2   + & \verb+~t2+ & \verb+~T2+ & \verb+0     + &
\verb+MStop2+ & \verb+wStop2+ & \verb+3    + & \\
\verb+b-squark 1   + & \verb+~b1+ & \verb+~B1+ & \verb+0     + &
\verb+MSbot1+ & \verb+wSbot1+ & \verb+3    + & \\
\verb+b-squark 2   + & \verb+~b2+ & \verb+~B2+ & \verb+0     + &
\verb+MSbot2+ & \verb+wSbot2+ & \verb+3    + & \\
\end{tabular}
\end{center}
\caption{MSSM particles table in the CompHEP notations.} 
\end{table}
}

The meaning of the table contents  is the following:
\bigskip

\begin{tabular}{ l l l}
\verb+Full name+ & -- & full particle name for the particular model;\\
\verb+P,AP     + & -- & particle and anti-particle notations;\\
\verb+2*spin   + & -- & doubled spin of a particle;\\
\verb+mass     + & -- & mass of a particle;\\
\verb+width    + & -- & width of a particle;\\
\verb+color    + & -- & transformation properties of a particle under the 
			$SU(3)_{colour}$ gauge group: \\
                 &    & 8 -- octet, 3 -- triplet and 1 -- singlet;\\
\verb+aux      + & -- & some specific properties of a particle.
\end{tabular}
\bigskip

The detailed explanation can be found in Ref.~\cite{comphep}

The MSSM contains, in general, many parameters:
\begin{itemize}
\item $SU(3)$, $SU(2)$ and $U(1)$ gauge couplings;
\item Yukawa couplings $h_L$, $h_U$, $h_D$ (they are, in general,
3-dimensional matrices in the generation space);
\item gaugino masses $M_1$, $M_2$, $M_3$;
\item trilinear soft supersymmetry breaking parameters $A_L$, $A_U$, $A_D$
(they are, in general, 3-di\-men\-si\-o\-nal matrices in the generation space);
\item Higgs mixing parameter $\mu$ and the corresponding bilinear soft
supersymmetry breaking parameter $B$ (the last can be re-expressed
through the ratio of vacuum expectation values of the Higgs fields 
$\tan\beta=v_2/v_1$);
\item a number of rotating matrices $Z_{ij}$ of squark, slepton, Higgs, 
chargino and neutralino sectors, as well as the CKM mixing matrix.
\end{itemize}

However, after the appropriate simplifying assumptions (unification of
the gauge coupling constants, universality of the soft supersymmetry
breaking parameters at the GUT scale, the diagonal form of the Yukawa
matrices, etc.) are made, only few independent parameters are left.
Below we discuss our assumptions and model parameters.

\begin{itemize}
\item  matrices of Yukawa couplings and corresponding trilinear soft 
supersymmetry breaking parameters $A$ are diagonal
\item superpartners of the left-handed and right-handed fermions of 
two light generations do not mix; mixing takes place only for the third 
generation sfermions. This turns to be rather accurate approximation, 
since the off-diagonal entries of the sfermion mass-squared-matrices are 
of the form $m_f(A_f-\mu\tan\beta)$. This assumption also fixes the form 
of the rotating matrices $Z_{ij}$ in squark and slepton sectors, for 
which we have also neglected the intergenerational mixing
\item we accept also some theoretical motivations, namely, the gauge 
coupling constant unification and the universality of the soft 
supersymmetry breaking parameters at the GUT scale, which is natural in, 
{\it e.g.} supergravity inspired models. This, however, affects only on 
numerical values of sparticle and Higgs masses and mixings which are 
presented below as an example. The last assumption can be relaxed and 
even rejected if one is interested in studying models beyond MSSM or 
effects of non-universal SUSY breaking terms, etc.
\end{itemize}

Under the above mentioned assumptions, in addition to the SM parameters 
one has the following set of parameters all taken at the Grand 
Unification scale (the corresponding model is often reffered to as 
Minimal Supergravity):
\begin{itemize}
\item $m_0$ -- the common mass of scalar particles
\item $m_{1/2}$ -- the common mass of fermions 
\item $\mu_0$ -- the initial value of the Higgs mixing parameter 
\item $A_0$ -- the initial value of trilinear soft supersymmetry breaking
parameters
\end{itemize}
The soft supersymmetry breaking part of the Lagrangian then takes the form:
$$
- L_{SB}  = m_0^2 \sum_{i}|\varphi_i|^2 
            + \left( m_{1/2}\sum_{\alpha} \lambda_\alpha \lambda_\alpha 
            + A_0 ( h_L \tilde L \tilde H_1 \tilde E^c
                  + h_D \tilde Q \tilde H_1 \tilde D^c
                  + h_U \tilde Q \tilde H_2 \tilde U^c)
            + B \mu \tilde H_1 \tilde H_2 + h.c. \right),
$$
$\varphi_i$ are the scalar particles, $\lambda_\alpha$ are gauginos, the 
tilde denotes the scalar component of the corresponding superfield, 
$SU(2)$ contraction being understood.
 
And the last parameter is
\begin{itemize}
\item $\tan\beta$ (the value of this parameter determines two different 
scenarios, the so called {\it high} and {\it low} $\tan\beta$ 
{\it scenarios}~\cite{deBoer})
\end{itemize}

The numerical values of the parameters can be chosen in different ways.
However, if one wants to perform a self-consistent analysis one has to 
be careful since many restrictions have to be satisfied simultaneously. 
We follow here the strategy of the global fit analysis~\cite{deBoer} in the 
framework of which one can predict values of parameters satisfying some 
common conditions and present experimental data. 

We use the following values of the input MSSM parameters obtained from 
the global fit analysis for high tan$\beta$ scenario (as an example) 
at the GUT scale:
\bigskip

\begin{center}
\begin{tabular}{ c | c | c | c | c | c | c | c | c | c }
$m_0$ & $m_{1/2}$ & $\mu$ & $\tan\beta$ & $Y_t$ & 
$Y_b$ & $Y_\tau$ & $M_{GUT}$ & $1/\alpha_{GUT}$ & $A_0$ \\ 
 & & & & & & & & & \\ \hline 
 & & & & & & & & & \\
800 & 88 & -270 & 41.2 & 0.0014 & 0.0011 & 0.0011 & 
$2.5\cdot 10^{16}$ & 24.3 & 0
\end{tabular}
\end{center}
\bigskip

\noindent
where $Y_i=h_i^2/16\pi^2$.

To calculate the numerical values of the soft SUSY breaking parameters 
and masses of superparticles we run one-loop renormalization group 
equations from the unification point down to the scale of the $Z$-boson 
mass. After that the values of the elements of rotating matrices 
$Z_{ij}$ can be also calculated.

The input parameters of the MSSM in the CompHEP notations are presented 
in Table~2 (the file \verb+vars+$N$\verb+.mdl+).

{\small
\begin{table}[htb]
\vspace*{-1.5cm}
\begin{tabular}{ l | l | l || l | l | l || l | l | l}
\verb+Name  +&\verb+Value    + &\verb+ Comment+ 	  &
\verb+Name  +&\verb+Value    + &\verb+ Comment+ 	  &
\verb+Name  +&\verb+Value    + &\verb+ Comment+ 	  \\
\verb+EE    +&\verb+ 0.31333 + & $\sqrt{4\pi\alpha_{em}}$ &
\verb+TB    +&\verb+ 41.2    + & $\tan\beta$		  &
\verb+wSe1  +&\verb+ 7.70    + & $\Gamma(\tilde e_1)$	  \\
\verb+GG    +&\verb+ 1.117   + & $\sqrt{4\pi\alpha_{s} }$ &
\verb+hx    +&\verb+-220     + & $\mu$  		  &
\verb+MSe2  +&\verb+ 806     + & $m_{\tilde e_2}$         \\
\verb+SW    +&\verb+ 0.474   + & $sin \theta_W$ 	  &
\verb+ls1   +&\verb+ 0       + & $A_L^1$		  &
\verb+wSe2  +&\verb+ 3.39    + & $\Gamma(\tilde e_2)$	  \\
\verb+s12   +&\verb+ 0.221   + & CKM			  &
\verb+ls2   +&\verb+ 0       + & $A_L^2$		  &
\verb+MSmu1 +&\verb+ 804     + & $m_{\tilde \mu_1}$	  \\
\verb+s23   +&\verb+ 0.04    + & CKM			  &
\verb+ls3   +&\verb+ 4.62    + & $A_\tau$		  &
\verb+wSmu1 +&\verb+ 7.70    + & $\Gamma(\tilde \mu_1)$   \\
\verb+s13   +&\verb+ 0.0035  + & CKM			  &
\verb+us1   +&\verb+ 0       + & $A_U^1$		  &
\verb+MSmu2 +&\verb+ 806     + & $m_{\tilde \mu_2}$	  \\
\verb+MZ    +&\verb+ 91.187  + & $M_Z$  		  &
\verb+us2   +&\verb+ 0       + & $A_U^2$		  &
\verb+wSmu2 +&\verb+ 3.39    + & $\Gamma(\tilde \mu_2)$   \\
\verb+Zn11  +&\verb+ 0.152   + & $Z_N^{11}$		  &
\verb+us3   +&\verb+ 106     + & $A_t^3$		  &
\verb+MStau1+&\verb+ 557     + & $m_{\tilde \tau_1}$	  \\
\verb+Zn12  +&\verb+ 0.060   + & $Z_N^{12}$		  &
\verb+ds1   +&\verb+ 0       + & $A_D^1$		  &
\verb+wStau1+&\verb+ 27.5    + & $\Gamma(\tilde \tau_1)$  \\
\verb+Zn13  +&\verb+-0.953   + & $Z_N^{13}$		  &
\verb+ds2   +&\verb+ 0       + & $A_D^2$		  &
\verb+MStau2+&\verb+ 695     + & $m_{\tilde \tau_2}$	  \\
\verb+Zn14  +&\verb+-0.255   + & $Z_N^{14}$		  &
\verb+ds3   +&\verb+ 3.32    + & $A_b$	                  &
\verb+wStau2+&\verb+ 36.5    + & $\Gamma(\tilde \tau_2)$  \\
\verb+Zn21  +&\verb+ 0.681   + & $Z_N^{21}$		  &
\verb+wZ    +&\verb+ 2.502   + & $\Gamma(Z)$		  &
\verb+MSne  +&\verb+ 801     + & $m_{\tilde \nu_e}$	  \\
\verb+Zn22  +&\verb+-0.172   + & $Z_N^{22}$		  &
\verb+wW    +&\verb+ 2.094   + & $\Gamma(W)$		  &
\verb+wSne  +&\verb+ 10.8    + & $\Gamma(\tilde \nu_e)$   \\
\verb+Zn23  +&\verb+ 0.274   + & $Z_N^{23}$		  &
\verb+Mm    +&\verb+ 0.1057  + & $m_{\mu}$		  &
\verb+MSnmu +&\verb+ 801     + & $m_{\tilde \nu_\mu}$	  \\
\verb+Zn24  +&\verb+-0.657   + & $Z_N^{24}$		  &
\verb+Mt    +&\verb+ 1.777   + & $m_{\tau}$		  &
\verb+wSnmu +&\verb+ 16.3    + & $\Gamma(\tilde \nu_\mu)$ \\
\verb+Zn31  +&\verb+-0.092   + & $Z_N^{31}$		  &
\verb+Mc    +&\verb+ 1.3     + & $m_{c}$		  &
\verb+MSntau+&\verb+ 801     + & $m_{\tilde \nu_\tau}$    \\
\verb+Zn32  +&\verb+-0.983   + & $Z_N^{32}$		  &
\verb+Ms    +&\verb+ 0.2     + & $m_{s}$		  &
\verb+wSntau+&\verb+ 35.4    + & $\Gamma(\tilde \nu_\tau)$\\
\verb+Zn33  +&\verb+-0.107   + & $Z_N^{33}$		  &
\verb+Mtop  +&\verb+ 175     + & $m_{t}$		  &
\verb+MSu1  +&\verb+ 831     + & $m_{\tilde u_1}$	  \\
\verb+Zn34  +&\verb+ 0.117   + & $Z_N^{34}$		  &
\verb+wtop  +&\verb+ 1.442   + & $\Gamma(t)$		  &
\verb+wSu1  +&\verb+ 55.8    + & $\Gamma(\tilde u_1)$	  \\
\verb+Zn41  +&\verb+ 0.710   + & $Z_N^{41}$		  &
\verb+Mb    +&\verb+ 4.3     + & $m_{b}$		  &
\verb+MSu2  +&\verb+ 830     + & $m_{\tilde u_2}$	  \\
\verb+Zn42  +&\verb+ 0.025   + & $Z_N^{44}$		  &
\verb+Mh    +&\verb+ 110     + & $m_{h}$		  &
\verb+wSu2  +&\verb+ 48.0    + & $\Gamma(\tilde u_2)$	  \\
\verb+Zn43  +&\verb+-0.072   + & $Z_N^{43}$		  &
\verb+wh    +&\verb+ 0.088   + & $\Gamma(h)$		  &
\verb+MSd1  +&\verb+ 835     + & $m_{\tilde d_1}$	  \\
\verb+Zn44  +&\verb+ 0.700   + & $Z_N^{44}$		  &
\verb+MHH   +&\verb+ 273     + & $m_{H}$		  &
\verb+wSd1  +&\verb+ 157.5   + & $\Gamma(\tilde d_1)$	  \\
\verb+Zm11  +&\verb+ 0.918   + & $Z_-^{11}$		  &
\verb+wHh   +&\verb+ 18.6    + & $\Gamma(H)$		  &
\verb+MSd2  +&\verb+ 831     + & $m_{\tilde d_2}$	  \\
\verb+Zm12  +&\verb+ 0.397   + & $Z_-^{12}$		  &
\verb+MH3   +&\verb+ 273     + & $m_{A}$		  &
\verb+wSd2  +&\verb+ 46.9    + & $\Gamma(\tilde d_2)$	  \\
\verb+Zm21  +&\verb+-0.397   + & $Z_-^{21}$		  &
\verb+wH3   +&\verb+ 18.8    + & $\Gamma(A)$		  &
\verb+MSc1  +&\verb+ 831     + & $m_{\tilde c_1}$	  \\
\verb+Zm22  +&\verb+ 0.918   + & $Z_+^{22}$		  &
\verb+MHc   +&\verb+ 285     + & $m_{H^\pm}$		  &
\verb+wSc1  +&\verb+ 55.8    + & $\Gamma(\tilde c_1)$	  \\
\verb+Zp11  +&\verb+ 0.994   + & $Z_+^{11}$		  &
\verb+wHc   +&\verb+ 9.73    + & $\Gamma(H^\pm)$	  &
\verb+MSc2  +&\verb+ 830     + & $m_{\tilde c_2}$	  \\
\verb+Zp12  +&\verb+ 0.106   + & $Z_+^{12}$		  &
\verb+MC1   +&\verb+ 65      + & $m_{\chi^\pm_1}$	  &
\verb+wSc2  +&\verb+ 48.0    + & $\Gamma(\tilde c_2)$	  \\
\verb+Zp21  +&\verb+-0.106   + & $Z_+^{21}$		  &
\verb+wC1   +&\verb+ 0.00003 + & $\Gamma(\chi^\pm_1)$	  &
\verb+MSs1  +&\verb+ 835     + & $m_{\tilde s_1}$	  \\
\verb+Zp22  +&\verb+ 0.994   + & $Z_+^{22}$		  &
\verb+MC2   +&\verb+ 254     + & $m_{\chi^\pm_1}$	  &
\verb+wSs1  +&\verb+ 57.5    + & $\Gamma(\tilde s_1)$	  \\
\verb+Zd33  +&\verb+-0.978   + & $Z_D^{33}$		  &
\verb+wC2   +&\verb+ 7.26    + & $\Gamma(\chi^\pm_2)$	  &
\verb+MSs2  +&\verb+ 831     + & $m_{\tilde s_2}$	  \\
\verb+Zd36  +&\verb+-0.206   + & $Z_D^{36}$		  &
\verb+MNE1  +&\verb+ 35      + & $m_{\chi^0_1}$ 	  &
\verb+wSs2  +&\verb+ 46.9    + & $\Gamma(\tilde s_2)$	  \\
\verb+Zd63  +&\verb+-0.206   + & $Z_D^{63}$		  &
\verb+wNE1  +&\verb+ 0       + & $\Gamma(\chi^0_1)$	  &
\verb+MStop1+&\verb+ 461     + & $m_{\tilde t_1}$	  \\
\verb+Zd66  +&\verb+ 0.978   + & $Z_D^{66}$		  &
\verb+MNE2  +&\verb+ 65      + & $m_{\chi^0_2}$ 	  &
\verb+wStop1+&\verb+ 16.0    + & $\Gamma(\tilde t_1)$	  \\
\verb+Zu33  +&\verb+ 0.158   + & $Z_U^{33}$		  &
\verb+wNE2  +&\verb+ 0	     + & $\Gamma(\chi^0_2)$	  &
\verb+MStop2+&\verb+ 543     + & $m_{\tilde t_2}$	  \\
\verb+Zu36  +&\verb+-0.987   + & $Z_U^{36}$		  &
\verb+MNE3  +&\verb+ 240     + & $m_{\chi^0_3}$ 	  &
\verb+wStop2+&\verb+ 39.7    + & $\Gamma(\tilde t_2)$	  \\
\verb+Zu63  +&\verb+ 0.987   + & $Z_U^{63}$		  &
\verb+wNE3  +&\verb+ 1.91    + & $\Gamma(\chi^0_3)$	  &
\verb+MSbot1+&\verb+ 463     + & $m_{\tilde b_1}$	  \\
\verb+Zu66  +&\verb+ 0.158   + & $Z_U^{66}$		  &
\verb+MNE4  +&\verb+ 248     + & $m_{\chi^0_4}$ 	  &
\verb+wSbot1+&\verb+ 29.8    + & $\Gamma(\tilde b_1)$	  \\
\verb+Zl33  +&\verb+-0.733   + & $Z_L^{33}$		  &
\verb+wNE4  +&\verb+ 11.5    + & $\Gamma(\chi^0_4)$	  &
\verb+MSbot2+&\verb+ 549     + & $m_{\tilde b_1}$	  \\
\verb+Zl36  +&\verb+-0.680   + & $Z_L^{36}$		  &
\verb+MSG   +&\verb+ 236     + & $m_{\tilde g}$ 	  &
\verb+wSbot2+&\verb+ 40.7    + & $\Gamma(\tilde b_2)$	  \\
\verb+Zl63  +&\verb+-0.680   + & $Z_L^{63}$		  &
\verb+wSG   +&\verb+ 0.000258+ & $\Gamma(\tilde g)$	  &
\verb+Maux  +&\verb+ 1       + & $m_{aux field}$	  \\		     
\verb+Zl66  +&\verb+ 0.733   + & $Z_L^{66}$		  &
\verb+MSe1  +&\verb+ 804     + & $m_{\tilde e_1}$	  &
             &                 &                          \\
\end{tabular}  
\caption{The input parameters of the MSSM in the CompHEP
notations.}
\end{table}
}

It should be stressed that widths of the particles have been calculated 
by means of CompHEP itself for this particular set of particle masses. 
One should keep the right widths of the particles for the calculation 
of different processes (especially, resonant ones) and recalculate 
widths for any new set of parameters.

The table of the MSSM Lagrangian (the file \verb+lgrng+$N$\verb+.mdl+)
exactly corresponds to the order of the Feynman rules in Ref.~\cite{rosiek}, 
but the section 14. All four-scalar vertices from section 14 and 
gluon-gluon-squark-squark vertices from section 15 are converted into 
three-particle vertices and given in the end of the table.

The four-scalar vertices originate from the scalar potential which is the 
sum of the $F$- and $D$-term parts:
$$ V=\frac{1}{2}(D^a_G D^a_G + D^a_W D^a_W + D^{}_B D^{}_B) 
+ F^*_i F^{}_i, $$
where
\begin{eqnarray*}
D^a_G & = & g_s(\tilde Q^{I*}\lambda^a\tilde Q^I+\tilde
D^{I*}\lambda^a\tilde D^I+\tilde U^{I*}\lambda^a\tilde U^I)
\\
D^a_W & = & \frac{g}{2}
(\tilde Q^{I*}\tau^a\tilde Q^I+\tilde L^{I*}\tau^a\tilde L^I
+H^{*}_1\tau^a H^{}_1+H^{*}_2\tau^a H^{}_2)
\\
D_B & = & g'
(\frac{1}{6}\tilde Q^{ I*}\tilde Q^I+\frac{1}{3}\tilde D^{ I*}
\tilde D^I-\frac{2}{3}\tilde U^{ I*}\tilde U^I-
\frac{1}{2}\tilde L^{ I*}\tilde L^I+\tilde E^{ I*}\tilde E^I
-\frac{1}{2} H^{*}_1 H^{}_1+\frac{1}{2} H^{*}_2 H^{}_2)
\\
F^*_{H_1} F^{}_{H_1}&=&\mu^2 H^*_2 H^{}_2 + 
h_L^I h_L^J \tilde L^{I*}\tilde L^J\tilde E^{I*}\tilde E^J+
h_D^I h_D^J \tilde Q^{''I*}\tilde Q^{''J}\tilde D^{I*}\tilde D^J\\
&&+ \left[ h_L^I \mu H_2^*\tilde L^I\tilde E^I+
h_D^I \mu H_2^*C^{IJ}\tilde Q^J\tilde D^I+
h_L^I h_D^J \tilde L^I C^{JK}\tilde Q^K\tilde E^I\tilde D^{*J}
+ H.c. \right]
\\
F^*_{H_2} F^{}_{H_2}&=& \mu^2 H^*_1 H^{}_1 + h_U^I h_U^J
\tilde Q^{''I*}\tilde Q^{''J}\tilde U^{I*}\tilde U^J -
\left[h_U^I \mu H_1^*\tilde Q^I\tilde U^I + H.c.\right]
\\
F^*_L F^{}_L&=&\left(h_L^I\right)^2 
H_1^*H^{}_1\tilde E^*_{I}\tilde E^{}_I
\\
F^*_E F^{}_E&=&\left(h_L^I\right)^2
\epsilon_{ij}\epsilon_{kl}
H_{1i}^* H^{}_{1k}\tilde L^{*I}_j\tilde L^I_l
\\
F^*_Q F^{}_Q&=&\left(h_D^I\right)^2 
H_1^*H^{}_1\tilde D^*_{I}\tilde D^{}_I+
 \left(h_U^I\right)^2 H_2^*H^{}_2\tilde U^*_{I}\tilde U^{}_I+
 \left[ h_U^I h_D^J H_1^*H^{}_2 C^{IJ} \tilde
U^{*I}\tilde D^J + H.c. \right]
\\
F^*_U F^{}_U&=&\left(\frac{m_u^I}{v_2}\right)^2\epsilon_{ij}\epsilon_{kl}
H_{2i}^*H^{}_{2k}\tilde Q^{'*I}_j\tilde Q^{'I}_l
\\
F^*_D F^{}_D&=&\left(\frac{m_d^I}{v_1}\right)^2\epsilon_{ij}\epsilon_{kl}
H_{1i}^*H^{}_{1k}\tilde Q^{''*I}_j\tilde Q^{''I}_l
\end{eqnarray*}

\noindent
$\lambda^a (a=1,\dots,8)$ and $\tau^a (a=1,2,3)$ are Gell-Mann and Pauli 
matrices, $C^{IJ}$ is the Cabbibo-Kobayashi-Maskawa mixing matrix, 
$ Q'^I_1 = Q^I_1, Q'^I_2=C^{IJ} Q^J_2 $ and
$ Q''^I_1 = C^{IJ} Q^J_1, Q''^I_2 = Q^I_2 $.

In order to reduce the enormous number of four-scalar vertices due to
flavour permutations we split each vertex to a pair of three-particle 
ones and introduce a number of auxiliary fields. The other reason of 
doing this is that 4-colour vertices cannot be implemented directly into 
CompHEP due to conventions about the colour structure. The way we 
introduced such vertices is clearly seen from the structure of 
$F$-terms which can be written as follows: 

\begin{eqnarray*}
F^*_{H_1}F^{}_{H_1}&=& |h_D^{IJ}\tilde Q^{''I}\tilde D^{*J}
+ h_L^{IJ}\tilde L^{I}\tilde E^{*J}+\mu \tilde H_2|^2
\\
F^*_{H_2}F^{}_{H_2}&=& |h_U^{IJ}\tilde Q^{'I}\tilde U^{*J}- \mu \tilde H_1|^2
\\
F^*_{L}F^{}_{L}&=& |h_L^{I}\tilde H_1 \tilde E^{*J}|^2
\\
F^*_{E}F^{}_{E}&=& |h_L^{I}\epsilon_{ij}\tilde H^{}_{1i}\tilde L^{*J}_j|^2
\\
F^*_{Q}F^{}_{Q}&=& |C^{*IJ}h_D^{JK}\tilde D^K \tilde H_1 
+ h_U^{IJ}\tilde U^J \tilde H_2|^2
\\
F^*_{U}F^{}_{U}&=& |h_U^{I}\epsilon_{ij}\tilde H^{}_{2i}\tilde Q^{'*J}_j|^2
\\
F^*_{D}F^{}_{D}&=& |h_D^{I}\epsilon_{ij}\tilde H^{}_{1i}\tilde Q^{''*J}_j|^2
\end{eqnarray*}

For example, the four-scalar vertices coming from terms $F_{H_1}^* F_{H_1}$ 
and $F_{H_2}^* F_{H_2}$ can be introduced through two doublets of the 
auxilary fields ($\xi_{1i}$, $\xi_{2i}$) with a constant propagator 
$1/M_\xi^2$. To cancel the dependence of the results on the mass of the 
auxiliary fields we multiply each vertex containing the latter by the 
factor $M_\xi$, however, it is necessary for CompHEP to define it and 
assign a numerical value for it (we put $M_\xi=1$).

These auxiliary fields are defined in the CompHEP particle table 
(the file \verb+prtcls+$N$\verb+.mdl+). The part of the 
file with the definition of new auxiliary fields is presented in Table~3.   

{\small
\begin{table}[htb]
\begin{center}
\begin{tabular}{ l | l | l | l | l | l | l | l }
\verb+imprt GGU1U1  + & \verb+~00+ & \verb+~01+ & \verb+2     + &
\verb+Maux  + & \verb+0    + & \verb+3    + & \verb+*+ \\
\verb+imprt GGU2U2  + & \verb+~02+ & \verb+~03+ & \verb+2     + &
\verb+Maux  + & \verb+0    + & \verb+3    + & \verb+*+ \\
\verb+imprt GGD1D1  + & \verb+~04+ & \verb+~05+ & \verb+2     + &
\verb+Maux  + & \verb+0    + & \verb+3    + & \verb+*+ \\
\verb+imprt GGD2D2  + & \verb+~06+ & \verb+~07+ & \verb+2     + &
\verb+Maux  + & \verb+0    + & \verb+3    + & \verb+*+ \\
\verb+imprt GGC1C1  + & \verb+~08+ & \verb+~09+ & \verb+2     + &
\verb+Maux  + & \verb+0    + & \verb+3    + & \verb+*+ \\
\verb+imprt GGC2C2  + & \verb+~0A+ & \verb+~0B+ & \verb+2     + &
\verb+Maux  + & \verb+0    + & \verb+3    + & \verb+*+ \\
\verb+imprt GGS1S1  + & \verb+~0C+ & \verb+~0D+ & \verb+2     + &
\verb+Maux  + & \verb+0    + & \verb+3    + & \verb+*+ \\
\verb+imprt GGS2S2  + & \verb+~0E+ & \verb+~0F+ & \verb+2     + &
\verb+Maux  + & \verb+0    + & \verb+3    + & \verb+*+ \\
\verb+imprt GGT1T1  + & \verb+~0G+ & \verb+~0H+ & \verb+2     + &
\verb+Maux  + & \verb+0    + & \verb+3    + & \verb+*+ \\
\verb+imprt GGT2T2  + & \verb+~0I+ & \verb+~0J+ & \verb+2     + &
\verb+Maux  + & \verb+0    + & \verb+3    + & \verb+*+ \\
\verb+imprt GGB1B1  + & \verb+~0K+ & \verb+~0L+ & \verb+2     + &
\verb+Maux  + & \verb+0    + & \verb+3    + & \verb+*+ \\
\verb+imprt GGB2B2  + & \verb+~0M+ & \verb+~0N+ & \verb+2     + &
\verb+Maux  + & \verb+0    + & \verb+3    + & \verb+*+ \\
\verb+imprt DD-SU3  + & \verb+~0O+ & \verb+~0O+ & \verb+0     + &
\verb+Maux  + & \verb+0    + & \verb+8    + & \verb+*+ \\
\verb+imprt SU2-1  + & \verb+~0P+ & \verb+~0P+ & \verb+0     + &
\verb+Maux  + & \verb+0    + & \verb+1    + & \verb+*+ \\
\verb+imprt SU2-2  + & \verb+~0Q+ & \verb+~0Q+ & \verb+0     + &
\verb+Maux  + & \verb+0    + & \verb+1    + & \verb+*+ \\
\verb+imprt SU2-3  + & \verb+~0R+ & \verb+~0R+ & \verb+0     + &
\verb+Maux  + & \verb+0    + & \verb+1    + & \verb+*+ \\
\verb+imprt U1     + & \verb+~0S+ & \verb+~0S+ & \verb+0     + &
\verb+Maux  + & \verb+0    + & \verb+1    + & \verb+*+ \\
\verb+imprt xi11  + & \verb+~0T+ & \verb+~0U+ & \verb+0     + &
\verb+Maux  + & \verb+0    + & \verb+1    + & \verb+*+ \\
\verb+imprt xi12  + & \verb+~0V+ & \verb+~0W+ & \verb+0     + &
\verb+Maux  + & \verb+0    + & \verb+1    + & \verb+*+ \\
\verb+imprt xi21  + & \verb+~0X+ & \verb+~0Y+ & \verb+0     + &
\verb+Maux  + & \verb+0    + & \verb+1    + & \verb+*+ \\
\verb+imprt xi22  + & \verb+~0Z+ & \verb+~0a+ & \verb+0     + &
\verb+Maux  + & \verb+0    + & \verb+1    + & \verb+*+ \\
\end{tabular}
\end{center}
\caption{The part of the CompHEP table of particles 
with the definition of new auxiliary fields.} 
\end{table}
}  

Some part of the CompHEP table of the MSSM Lagrangian with auxiliary
fields is presented in Table~4 as an example (the vertices containing
sparticles of the first and second generation, and conjugated vertices
are skipped).

{\small
\begin{table}[htb]
\vspace*{-1.5cm}
\begin{center}
\begin{tabular}{ l| l| l| l| l| l}
\verb+P1+ & \verb+P2+ & \verb+P3+ & \verb+P4+ & \verb+Factor+ &
\verb+Lorentz part+ \\
\verb+G+ & \verb+~0N+ & \verb+~b2+ &    & \verb+Sqrt2*GG*Maux+ & \verb+m1.m2+ \\
\verb+G+ & \verb+~0M+ & \verb+~B2+ &    & \verb+Sqrt2*GG*Maux+ & \verb+m1.m2+ \\
\verb+G+ & \verb+~0L+ & \verb+~b1+ &    & \verb+Sqrt2*GG*Maux+ & \verb+m1.m2+ \\
\verb+G+ & \verb+~0K+ & \verb+~B1+ &    & \verb+Sqrt2*GG*Maux+ & \verb+m1.m2+ \\
\verb+G+ & \verb+~0J+ & \verb+~t2+ &    & \verb+Sqrt2*GG*Maux+ & \verb+m1.m2+ \\
\verb+G+ & \verb+~0I+ & \verb+~T2+ &    & \verb+Sqrt2*GG*Maux+ & \verb+m1.m2+ \\
\verb+G+ & \verb+~0H+ & \verb+~t1+ &    & \verb+Sqrt2*GG*Maux+ & \verb+m1.m2+ \\
\verb+G+ & \verb+~0G+ & \verb+~T1+ &    & \verb+Sqrt2*GG*Maux+ & \verb+m1.m2+ \\
\verb+~0O+ & \verb+~B1+ & \verb+~b1+ &    & \verb+i*2*GG/2*Maux+       
& \verb+1+                 \\
\verb+~0O+ & \verb+~T1+ & \verb+~t1+ &    & \verb+i*2*GG/2*Maux+       
& \verb+1+                 \\
\verb+~0R+ & \verb+~E6+ & \verb+~e6+ &    & \verb+-i*2*EE/(4*SW)*Maux+ 
& \verb+ZL36**2+           \\
\verb+~0Q+ & \verb+~E6+ & \verb+~n3+ &    & \verb+-2*EE/(4*SW)*Maux+   
& \verb+ZL36+              \\
\verb+~0P+ & \verb+~E6+ & \verb+~n3+ &    & \verb+i*2*EE/(4*SW)*Maux+  
& \verb+ZL361+             \\
\verb+~0R+ & \verb+~N3+ & \verb+~n3+ &    & \verb+i*2*EE/(4*SW)*Maux+  
& \verb+1+                 \\
\verb+~0R+ & \verb+~T1+ & \verb+~t1+ &    & \verb+i*2*EE/(4*SW)*Maux+  
& \verb+ZU33**2+           \\
\verb+~0Q+ & \verb+~C1+ & \verb+~b1+ &    & \verb+2*EE/(4*SW)*Maux+    
& \verb+Vcb*ZD33*ZU22+     \\
\verb+~0P+ & \verb+~C1+ & \verb+~b1+ &    & \verb+i*2*EE/(4*SW)*Maux+  
& \verb+Vcb*ZD33*ZU22+     \\
\verb+~0S+ & \verb+~E6+ & \verb+~e6+ &    & \verb+i*2*EE/(4*CW)*Maux+  
& \verb+2*ZL66**2-ZL36**2+ \\
\verb+~0S+ & \verb+~N3+ & \verb+~n3+ &    & \verb+-i*2*EE/(4*CW)*Maux+ 
& \verb+1+                 \\
\verb+~0S+ & \verb+~B2+ & \verb+~b2+ &    & \verb+i*2*EE/(12*CW)*Maux+ 
& \verb=2*ZD66**2+ZD36**2= \\
\verb+~0S+ & \verb+~T2+ & \verb+~t2+ &    & \verb+-i*2*EE/(12*CW)*Maux+
& \verb+4*ZU66**2-ZU36**2+ \\
\verb+~0W+ & \verb+~E6+ & \verb+~e6+ &    & \verb+-Maux+ &
\verb+ZL36*ZL66*l3+     \\
\verb+~0U+ & \verb+~E3+ & \verb+~n3+ &    & \verb+-Maux+ &
\verb+ZL63*l3+          \\
\verb+~0W+ & \verb+~B1+ & \verb+~b1+ &    & \verb+-Maux+ &
\verb+ZD33*ZD63*d3+     \\
\verb+~0U+ & \verb+~B1+ & \verb+~t1+ &    & \verb+-Maux+ &
\verb+Vtb*ZD63*ZU33*d3+ \\
\verb+~0V+ & \verb+~E6+ & \verb+~e6+ &    & \verb+Maux+  &
\verb+ZL36*ZL66*l3+     \\
\verb+~0T+ & \verb+~N3+ & \verb+~e3+ &    & \verb+Maux+  &
\verb+ZL63*l3+	   \\
\verb+~0V+ & \verb+~B2+ & \verb+~b2+ &    & \verb+Maux+  &
\verb+ZD36*ZD66*d3+     \\
\verb+~0T+ & \verb+~T2+ & \verb+~b2+ &    & \verb+Maux+  &
\verb+Vtb*ZD66*ZU36*d3+ \\
\verb+~0a+ & \verb+~T1+ & \verb+~b1+ &    & \verb+-Maux+ &
\verb+Vtb*ZD33*ZU63*u3+ \\
\verb+~0Y+ & \verb+~T2+ & \verb+~t2+ &    & \verb+-Maux+ &
\verb+ZU36*ZU66*u3+     \\
\verb+~0Z+ & \verb+~B2+ & \verb+~t2+ &    & \verb+Maux+  &
\verb+Vtb*ZD36*ZU66*u3+ \\
\verb+~0X+ & \verb+~T2+ & \verb+~t2+ &    & \verb+Maux+  &
\verb+ZU36*ZU66*u3+     \\
\end{tabular}
\end{center}
\caption{The part of the CompHEP table of the MSSM Lagrangian with auxiliary
fields}
\end{table}
}

\newpage
%
%
\section{Test of the model and conclusions}

In the paper we have presented the Minimal Supersymmetric Standard Model 
as a model implemented into the CompHEP package. The model has been already
used for the study of the chargino pair production at  
LEP~\cite{chargino-lep}. Chargino sector has been tested by comparison of 
the analytical results obtained by means of CompHEP with those
presented in~\cite{chargino1,chargino2}. In Ref.~\cite{chargino1} the chargino 
pair production at high energy $\gamma\gamma$ colliders has been studied 
($\gamma\gamma\rightarrow\tilde\chi^-_j \tilde\chi^+_j$ process).
The paper~\cite{chargino2} is devoted to the study of chargino and sneutrino 
production in electron-photon collisions
($e^-\gamma\rightarrow\tilde\chi^-_j \tilde\nu_e$ process).
CompHEP results are in agreement with the results obtained in these papers.
The charged Higgs sector has been also tested. We have an agreement with the
results of Ref.~\cite{chiggs} where the study of the charged Higgs pair 
production in $e^+ e^-$ collisions has been performed. 
\bigskip

People from High Energy Physics community are welcome to study
the MSSM within the CompHEP software package. Any remarks or suggestions are 
appreciated.

%
%
\subsection*{Acknowledgments}

We would like to thank S.Ambrosanio, E.Boos, V.Ilyin, D.Kazakov, D.P.Roy
for valuable discussions and their help in implementing the MSSM into the 
CompHEP package. Espesially must we express our gratitude to A.Pukhov who 
provided us the new improved version of the CompHEP package which is more 
effective for calculations in the framework of complicated models and MSSM 
in particular.

The financial support of the Russian Foundation for Basic
Research  (grants \# 96-02-17379-a, \# 96-02-19773-a)  and ICFPM in 1996 is 
acknowledged. The work of A.V.S. was also supported by the grants 
ISSEP a97-966 and INTAS 93-1180-ext.

\clearpage

%
%

\end{document}